\documentclass{eas}
\usepackage{graphicx}
\usepackage{natbib}
\usepackage{journaux}

\begin{document}

\title{Stellar Dynamics of low mass stars from the surface to the interior measured by CoRoT and \emph{Kepler}} 
\runningtitle{Stellar Dynamics of low mass stars}
\author{Rafael A. Garc\'\i a} 
\address{IRFU, CEA, Universit\'e Paris-Saclay, F-91191 Gif-sur-Yvette, France}
\address{Universit\'e Paris Diderot, AIM, Sorbonne Paris Cit\'e, CEA, CNRS, F-91191 Gif-sur-Yvette, France}
%
%
\begin{abstract}
Continuous high-precision photometry of stars --provided by space missions such as CoRoT, \emph{Kepler}, and K2-- represents a unique way to study stellar rotation and magnetism. The coupling of these studies of the surface dynamics with asteroseismology is changing our view to surface and internal dynamics. In this proceedings I will provide the latest developments in the understanding  of surface and internal rotation and magnetic fields. I will also discuss the possible discovery of strong internal magnetic fields of dynamo origin in the convective cores of stars above 1.2-1.4 solar masses. I will finish by providing constraints on gyrochronology laws for low-mass stars and put the Sun into context of its magnetism when compared to other solar-analog stars.

\end{abstract}
\maketitle
\section{Introduction}
Current photometric observations obtained from space missions such as CoRoT \citep{2006cosp...36.3749B}, \emph{Kepler} \citep{2010Sci...327..977B}, and K2 \citep{2014PASP..126..398H} allow us to monitor stellar brightness in a continuous way with very high precision for hundred of thousand stars \citep[e.g.][]{2016cole.book..109C,2017ApJS..229...30M}. These long time series, typically from 1 month to several years, have revolutionized the study of surface stellar dynamics. For stars exhibiting magnetic features on their surfaces (e.g. starspots), the stellar brightness is modulated with a period proportional to the stellar rotation rate at the latitude in which these spots are located \citep[e.g.][]{2009A&A...506...41G,2012ASPC..462..133G,2014A&A...572A..34G,2013A&A...557L..10N,2013ApJ...775L..11M,2014ApJS..211...24M,2016MNRAS.456..119C,2017A&A...602A..63B,2017A&A...605A.111C}. Therefore, in these stars, it is possible to measure the average surface rotation rate as a function of time. Moreover, if the star has a surface differential rotation and there are several spots evolving at different latitudes, it is possible to measure a lower limit of this differential rotation \citep[e.g.][]{2012A&A...543A.146F,2013A&A...557A..11R,2014A&A...564A..50L,2015A&A...583A..65R,2017A&A...599A...1S}. 

If there is a magnetic cycle in a star (regular or not), the size and number of spots at the surface changes inducing a long term evolution of the stellar surface brightness that can be used as a proxy of the surface magnetism \citep[e.g.][]{2009A&A...501..703O,2013ApJ...769...37B,2013A&A...550A..32M,2017A&A...608A..87S}. To ensure that the brightness variations are related to the magnetic evolution and not to other sources such as convection or pulsations, it has been proposed to sample such variations in chunks of three to five times the average stellar rotation rate \citep{2014JSWSC...4A..15M}. The corresponding magnetic proxy has been called photospheric S index, $S_{ph}$, in analogy to the Chromospheric S index \citep{1978ApJ...226..379W,1984ApJ...279..763N} and has been used to study stellar magnetism of different type of stars from F to M dwarfs \citep[e.g.][]{2014A&A...562A.124M,2014JSWSC...4A..15M,2014MNRAS.441.2744V,2016A&A...589A.118S,2016A&A...596A..31S}.

Long photometric space observations have also enabled the asteroseismic revolution. Asteroseismology has provided global stellar parameters such as masses and radii for thousands of stars from the direct characterization of the seismic properties 
\citep[e.g.][]{2014ApJS..210....1C,2017ApJS..233...23S,2017ApJ...835..173S}, or by indirect methods from the study of the convective properties \citep{2011ApJ...741..119M,2014A&A...570A..41K} or the study of the variability \citep{2013Natur.500..427B,2016ApJ...818...43B,2017arXiv171102890B,2018BugnetAA}. Moreover, today it is possible to probe stellar interiors and constrain their structures \citep[e.g.][]{2010A&A...518A..53M,2010ApJ...723.1583M,2011ApJ...733...95M,2011A&A...535A..91D,2012ApJ...749..152M,2012ApJ...748L..10M,2014ApJS..214...27M,2017A&A...601A..67C,2017ApJ...835..173S} and dynamic properties \citep[e.g.][]{2012ApJ...756...19D,2014A&A...564A..27D,2012A&A...548A..10M,2013PNAS..11013267G,2014A&A...563A..84G,2015MNRAS.452.2654B,2015A&A...582A..10N,2017A&A...603A...6N,2016ApJ...817...65D} for hundreds of main-sequence dwarfs and tens of thousands of red giants. The temporal variation of seismic parameters have also allowed us to unveil the presence of magnetic activity cycles in several main-sequence dwarfs \citep[e.g.][]{2010Sci...329.1032G,2013A&A...550A..32M,2016A&A...589A.103R,2016A&A...589A.118S,2017A&A...598A..77K,2018arXiv180600136S}.

In this proceedings the latest results concerning surface and internal rotation and magnetism combining photometric variations and asteroseismology are reviewed. Special attention has been given to update the bibliography to the moment in which this proceedings has been written.

\section{Stellar rotation}
\subsection{Surface}
Using \emph{Kepler} data, \cite{2013A&A...557L..10N},  \cite{2013A&A...560A...4R}, and \cite{2014ApJS..211...24M} have determined surface rotations, $P_{\rm{rot}}$, of tens of thousands of stars. All of them used Simple Aperture Photometry (SAP) time series, calibrated using PDC (Pre-search Data Conditioning) algorithms \citep{2012PASP..124..985S,2012PASP..124.1000S,2014PASP..126..100S}. Originally developed to maximize the exoplanet detections, PDC could filter stellar periods longer than 3 days  \citep[see for example the discussion in][]{2013ASPC..479..129G}. Therefore, it is recommended to use other calibration procedures ensuring that the low-frequency stellar signals including rotation are not filtered out. This is the objective of the KADACS light curves \citep{2011MNRAS.414L...6G}, which also include an effective interpolation procedure based on in-painting techniques \citep{2015A&A...574A..18P} to reduce the effects of the regular gaps in the \emph{Kepler} light curves \citep{2014A&A...568A..10G}, or the newly developed methods based on gaussian processes \citep{2017MNRAS.471..759A}.

Surface rotation can be extracted by selecting the highest peak in the low-frequency part of the power spectrum \citep[e.g.][]{2009A&A...506...51B,2011A&A...534A...6C,2013A&A...557L..10N}. When a star has two spots at approximately $180^o$, the highest peak is the second harmonic and the retrieved peak is half of the average $P_{\rm{rot}}$ \citep[see an extensive discussion in][]{2013ASPC..479..129G}. To minimize this problem,  it is recommended to study the rotation in the time domain through the use of autocorrelation functions \citep[e.g.][]{2013MNRAS.432.1203M}. This methodology is more robust and it is less sensitive to the above mentioned effect of measuring half rotation periods. Finally, it is possible to study rotation in a  time-period diagram based on wavelet transforms \citep[e.g.][]{2009A&A...506...41G,2010A&A...518A..53M}. This method has the advantage of highlighting possible instrumental problems in a given quarter. In a benchmark comparaison of different groups and techniques, \cite{2015MNRAS.450.3211A} showed that the best combination of completeness and reliability $P_{\rm{rot}}$ were obtained by combining different detrending algorithms and several period-search methods.

\subsection{Calibrating gyrochronology: contribution of asteroseismology}

One of the fundamental properties of stars is their ages. But ages are not easy to determine and the classical isochrone method \citep{1962ApJ...135..349S,1964ApJ...140..544D} is not precise enough to impose strict constraints to modern astrophysical problems, for example, to accurately date extrasolar systems of field stars. With the discovery of the evolution of chromospheric activity in cool stars \citep{1978ApJ...226..379W,1972ApJ...171..565S,1984ApJ...287..769N} activity-age relations developed. However, chromospheric activity can change in short time scales due to activity cycles, rotation phases, etc, limiting the precision in the age determination to around 50$\%$ \citep[see for example][]{2007ApJ...669.1167B}. Thus, rotation-age relations were proposed by \cite{1972ApJ...171..565S} using averaged $v \sin(i)$ and known ages, leading to gyrochronology, i.e., the determination of the age from the measurement of the stellar rotation period in cool main-sequence dwarfs. Having an external convective region, these stars lose angular momentum with time through stellar winds. The measure of the rotation rate is then an indication of their ages. The use of the rotation period instead of the projected rotation velocity allows one to avoid the dependency on the stellar inclination angle.  

To properly calibrate gyrochronology, stars belonging to young and intermediate-age clusters ($< 2.5$ Gyr) and the Sun are commonly used \citep[see for example][]{2015Natur.517..589M}. Thanks to space photometry, this has been extended to old open clusters such as M67 \citep{2016ApJ...823...16B}. However, these rotation periods need to be taken with caution as ``the sensibility drops rapidly with increasing period and decreasing amplitude, maxing at 15$\%$ recovery rate for the solar case (i.e. 25 d period, 0.1$\%$ amplitude)'' \citep[as discussed by ][]{2018arXiv180509922E}. Another interesting calibration set for old field stars is the seismic sample for which stellar ages are accurately computed through asteroseismic techniques \citep{2014A&A...572A..34G,2015MNRAS.450.1787A,2015MNRAS.446.2959D}. Field stars of the same mass or above the Sun have an anomalous fast rotation compared to the general predictions of the gyrochronology laws. Although still debated, \cite{2016Natur.529..181V} explained these discrepancy by a weakened magnetic braking. This result might suggest a fundamental change in the nature of evolved stellar dynamos, with the Sun being close to the critical transition to much weaker magnetized winds that occurs to all middle-aged stars \citep{2017SoPh..292..126M}. If confirmed, the diagnostic power of gyrochronology would be limited to stars younger than halfway their main-sequence lifetimes.

\subsection{Internal}
One of the main successes of helioseismology has been the inference of the solar internal differential rotation \citep[e.g.][]{ThoToo1996,1999MNRAS.308..405C,ThoJCD2003,2004SoPh..220..269G} including hints of the inner average core spin rate \citep{2007Sci...316.1591G,2017A&A...604A..40F}. When extending this analysis to other stars, the precision and the extension of the region to be explored depends on our ability to probe the inner regions, through mixed modes \citep{2011Sci...332..205B} in subgiants and red giants, or only the external regions, in main-sequence stars.

To have a general overview of the internal rotation rate of solar-like stars through evolution we start by main-sequence solar-like dwarfs. In these stars, neither mixed modes nor g modes have been measured. The only stratified information that can be obtained is through the careful analysis of acoustic modes. Unfortunately, although low-degree p modes penetrate deep in the stellar interior, they spend a small fraction of their time in the deep interior and the amount of information they can provide from these regions is small. 

\cite{2015MNRAS.452.2654B} analyzed CoRoT and \emph{Kepler} cool dwarfs and computed rotation splittings for $\ell$=1 and 2 modes and the stellar inclination angle. Assuming that the stellar interior is composed of a radiative and a convective region rotating uniformily and using surface rotation rates deduced from spectroscopy or directly extracted from the \emph{Kepler} photometry \citep{2014A&A...572A..34G} as a good indication of the average rotation rate of the external convective region \citep[as in the solar case, see also the discussion in][for the 16Cyg. system]{2015MNRAS.446.2959D}, they obtained an estimation of the rotation rate in the internal radiative region. They found that for all the stars in the sample but one, the gradient between the two zones is less than 2, which implies a rather flat equatorial rotation rate similar to the equatorial solar rotation profile \citep[see for example Fig.~9 in][]{2008SoPh..251..119G}. Moreover, because the internal structure of similar solar-like stars is comparable, a large number of those comparable stars could be use simultaneously to constrain their average radial differential rotation gradient in an ensemble way \citep{2016A&A...586A..79S}.

Leaving the main sequence \citep{2012ApJ...756...19D,2014A&A...564A..27D}, studied the internal rotation in subgiants and young red giants. To do so, they first characterized the rotational splittings of mixed modes in KIC~7341231. A clear gradient where found with larger splittings for modes with a larger g-mode contribution. This implied a faster core rotation when compared with the external regions. This analysis was then extended to 6 more stars \citep{2014A&A...564A..27D} obtaining similar results with a faster rotation rate in the core than in the surface. When using $\log g$ as a proxy of age, the core is contracting, the internal rotation is smoothly spinning up while the surface rotation is gradually slowing down as a consequence of the expansion of the external layers. This result is predicted by theory.

In the red giant phase, the first measurements of the gradient in RGB stars showed larger values than in main sequence or during the subgiant phase \citep[e.g.][]{2012Natur.481...55B,2016ApJ...817...65D}. However, the spinning rate is too low when compared with theoretical expectations \citep[e.g.][]{2013A&A...555A..54C} and a revision of angular momentum transport mechanisms is required \citep[see discussions in, e.g.][]{2013A&A...549A..74M,2013LNP...865...23M,2013ApJ...775L...1T,2017A&A...599A..18E}. Moreover, the evolution of the core rotation rate is different to what it was expected from theory. Thus, \cite{2012A&A...548A..10M} showed that the core rotation spins down when stars evolve in the RGB, i.e., when the radius increase for a given stellar mass. When stars reach the clump, the core rotation is reduced again. This can be partially explained by the expansion of the non-degenerate helium burning core \citep{1971PASP...83..697I,2000ApJ...540..489S}, but it is required a significant transfer of internal angular momentum from the inner to the outer layers in this evolutionary phase. A detailed analysis of seven second clump stars revealed a core-to-envelope gradient in a range 1.8 to 3.2 \citep{2015A&A...580A..96D}. However, the latest results of the core-rotation rates by \citep{2018arXiv180204558G} suggest that instead of a slight spin down of the core during the RGB phase, the core rotation seems to be constant independently of the mass. 

\section{Stellar magnetism}
\subsection{Surface}
Solar-like pulsating stars are characterized by having an external convective zone that excites oscillation modes. Convection and rotation usually generate dynamo magnetic fields in the surface of these stars that can then be measured and characterized. 

Using directly the photometric time series obtained from CoRoT and \emph{Kepler}, it is possible to obtain indications of the variability of the stars at different time scales \citep{2013ApJ...769...37B}. To ensure a magnetic origin of this variability, \cite{2014JSWSC...4A..15M} demonstrated that by taking chunks of 3-5 times the rotation period, the average resultant variability is a good proxy of the surface magnetism. This $S_{\rm{ph}}$ photometric activity proxy was recently applied to the Sun \citep{2017A&A...608A..87S}  using Sun-as-a-star VIRGO/SPM photometric light curves as well as Doppler velocity GOLF time series \citep{GarSTC2005}. The comparison with different solar magnetic proxies showed hysteresis patterns between all the proxies \citep[e.g.][]{1994SoPh..150..347B,1998A&A...329.1119J,2001SoPh..200....3T}. The hysteresis effects are a consequence of the different response to the magnetic perturbations at different heights of the solar atmosphere, which could also experience a temporal lag. This hysteresis pattern between chromospheric and photospheric proxies is also visible in other stars such as the young solar analog KIC~10644253 \citep{2016A&A...596A..31S}. 

Using an average $S_{\rm{ph}}$ for each solar-like pulsating dwarf observed by \emph{Kepler} during the survey phase \citep{2014ApJS..210....1C}, \cite{2014A&A...572A..34G} demonstrated that the surface magnetism of this ensemble set \citep{2011Sci...332..213C} is comparable to the one of the Sun. A detailed analysis of the photospheric and chromospheric activity of eighteen solar-analogs confirmed this result \citep{2016A&A...596A..31S}. 

Stars with larger surface magnetic variability show either weak or no oscillation modes \citep{2009A&A...506...33M,2011ApJ...732L...5C}, which has an impact on the prediction of stellar oscillations \citep{2011ApJ...732...54C}. Moreover, the amplitude of the oscillation modes is anticorrelated with the magnetic activity cycle as it was already measured in the Sun and in the CoRoT target HD~49933 \citep{2010Sci...329.1032G} and later in several \emph{Kepler} stars \citep{2017A&A...598A..77K}. However, according to the analyses done by \cite{2018arXiv180600136S}, 20$\%$ of their \emph{Kepler} sample present a positive correlation between the temporal evolution of the mode heights and the frequency shifts.

\subsection{Internal}
While the variation of the averaged frequency shifts with time provide information about possible on-going magnetic activity cycles \citep[e.g.][]{2017A&A...598A..77K,2018A&A...611A..84S,2018arXiv180600136S}, 
the frequency dependence of the p-mode frequency shifts can provide information about the structural changes occurring in stellar interiors with magnetic activity for solar-like dwarfs. In the case of the Sun, these changes are located in a thin layer right beneath the solar photosphere \citep[e.g.][]{1990Natur.345..779L,2016LRSP...13....2B}. More precisely, \cite{2012ApJ...758...43B} suggested that in cycle 23 those changes were localized mainly in a region above about 0.996 $R_{\odot}$. Using longer time series, \cite{2018arXiv180505298J} have showed that structural and magnetic changes responsible for the frequency shifts remained comparable between the last two cycles (23 and 24), while they are different from what was found in cycle 22.

This linear dependence of the frequency shifts with frequency was also found in other stars: first in the CoRoT target HD~49933 \citep{2011A&A...530A.127S} and later in the young solar analog KIC10644253 \citep{2016A&A...589A.118S}. In both cases, the perturbation in the acoustic modes was found to be located close to the photosphere. However, in four stars of the \emph{Kepler} sample analyzed by \cite{2018A&A...611A..84S}, KIC~5184732, KIC~8006161, KIC~8379927, and KIC~11081729, the frequency shifts normalized by the mode inertia show an oscillatory behavior instead of a linear one. This suggests that the frequency perturbation could be produced in a thin layer inside the resonant cavity of the acoustic modes because in that case, this frequency dependence of the frequency shifts would follow a sinusoidal function. Before driving firm conclusions about the positions and mechanisms responsible for the frequency shifts in stars, a larger stellar sample is required. However, we can already say today that the picture seems to be much more complicated than what it was outlined from the unique analysis of the Sun. 

The asteroseismic analysis of thousands of red giants has not yet reported the measure of any frequency shifts in the acoustic modes. This is probably due to a very small magnetic activity in the external layers of these stars. The same can be said about the study of mixed modes. Up to now, neither frequency shifts nor magnetic splittings have been reported. 

Things changed in 2011, at the 4th \emph{Kepler} Asteroseismic Science Consortium (KASC-4) meeting, when Garc\'ia et al. showed two red-giant stars observed by \emph{Kepler} in which the amplitudes of the dipole modes were unexpectedly small. Then, \cite{2012A&A...537A..30M}  realized that about 5$\%$ of their sample of 800 red giants had low-visibility dipolar modes. \cite{Fuller23102015} outlined an explanation through the presence of a strong magnetic field in the core of these giants. They called this mechanism the ``magnetic green house effect'' because the magnetic field acts as a green house trapping the energy of the modes propagating in the core \citep[see also][]{2016ApJ...824...14C}. Later, \cite{2016Natur.529..364S} suggested that such magnetic field could be of dynamo origin because after analyzing 13,000 red giants there were no depressed dipolar modes for masses below 1.2 to 1.3 $M_{\odot}$. In other words, only stars developing a convective core during their main-sequence life are affected, reaching about half of the stars for masses around 1.6  $M_{\odot}$. Although this explanation could solve the problem of the origin of magnetic white dwarfs (which are the cores of these red giants), the result is still controversial. Indeed, this was explained as a normal mixed-mode pattern while the green house effect mechanism expects full suppression of this mixed-mode pattern. Unfortunately, \cite{2017A&A...598A..62M} did not provide any further explanation of the suppression of the power. Nearly at the same time \cite{2017MNRAS.467.3212L} published another mechanism in which the magnetic field could still be responsible for the reduction in power of the dipolar modes while still allowing to have some mixed-mode energy. The last word is not said yet in this topic and the community is actively studying these stars to better understand the mechanism responsible for the reduction in power of these dipolar modes.

\section{Conclusions}
The high-precision photometry of thousands of stars from space has put stellar dynamics in the center of many important studies. In the last few years there has been a lot of observational new discoveries allowing us to develop and refine theoretical models. This not only affects stellar physics but also the Sun and the stellar magnetic evolution that can change the way in which our star could be conceived in the future if it is transiting to a more quite magnetic phase as lately suggested.

K2 and TESS missions, as well as the future ESA M3 PLATO mission \citep{2014ExA....38..249R} will provide a wealth of data. The future cannot be more promising for this field of research.

\section{Acknowledgements}
Funding for the \emph{Kepler} Discovery mission is provided by National Aeronautics and Space Administrations's Science Mission Directorate. The authors wish to thank the entire \emph{Kepler} team, without whom these results would not have been obtained. The CoRoT space mission has been developed and is operated by Centre National d'Etudes Spatiales (CNES), with contributions from Austria, Belgium, Brazil, the European Space Agency (RSSD and Science Program), Germany and Spain. The author thanks the French Programme National de Physique Stellaires for support as well as the CNES for supporting CoRoT and GOLF/SoHO activities at the DAp, CEA-Saclay. The author also acknowledges the support of the European Community's Seventh Framework Programme (FP7/2007-2013) under grant agreement No. 269194 (IRSES/- ASK) and No. 312844 (SPACEINN).

\bibliographystyle{aa}
\bibliography{./BIBLIO}
\end{document}